# The Nonequilibrium Mechanism of Noise-Enhanced Drug Synergy in HIV Latency Reactivation

Xiaolu Guo[1], Tao Tang[2], Minxuan Duan[3], Lei Zhang[4,5,*], Hao Ge[4,6,*]

[1]School of Mathematical Sciences, Peking University, Beijing, 100871, China
[2]School of Life Sciences, Peking University, Beijing, 100871, China
[3]Yuanpei College, Peking University, Beijing, 100871, China
[4]Beijing International Center for Mathematical Research, Peking University, Beijing, 100871, China
[5]Center for Quantitative Biology, Peking University, Beijing, 100871, China
[6]Biomedical Pioneering Innovation Center, Peking University, Beijing, 100871, China

*Correspondence: zhangl@math.pku.edu.cn (L.Z.), haoge@pku.edu.cn (H.G.)

## Abstract

Noise-modulating chemicals can synergize with transcriptional activators in reactivating latent HIV to eliminate latent HIV reservoirs. To understand the underlying biomolecular mechanism, we investigate a previous two-gene-state model and identify two necessary conditions for the synergy: an assumption of inhibition effect of transcription activators on noise enhancers; and frequent transitions to the gene non-transcription-permissive state. We then develop a loop-four-gene-state model with Tat transcription/translation and find that drug synergy is mainly determined by the magnitude and direction of energy input into the genetic regulatory kinetics of the HIV promoter. The inhibition effect of transcription activators is actually a phenomenon of energy dissipation in the nonequilibrium gene transition system. Overall, the loop-four-state model demonstrates that energy dissipation plays a crucial role in HIV latency reactivation, which might be useful for improving drug effects and identifying other synergies on lentivirus latency reactivation.

## Statement of Significance

By eliminating latent HIV reservoirs, the main barrier to a clinical cure, the “kick and kill” strategy has become a promising way to cure HIV. So far, two categories of biomolecules, activators (AC) and noise enhancers (NE), have been found to have synergy in reactivating latent HIV (the “kick” strategy). We uncover the underlying nonequilibrium mechanism of such a drug synergy by developing mathematical models based on genetic regulatory kinetics. We find that controlling the magnitude and direction of energy input into genetic regulatory kinetics can prevent NE from reducing the turn-on rate of the inactivated gene state in the presence of AC, which produces the synergy. This general nonequilibrium mechanism can be useful for identifying other drug synergies on lentivirus latency reactivation.

## Introduction

At the end of 2017, more than 36 million people were estimated to be infected with HIV (1). After HIV infects CD4+ cells, it can replicate or enter proviral latency (Figure 1A). Latent HIV reservoirs are the main obstacle to achieving a clinical cure (2). Reactivating latent HIV, quickly followed by antiretroviral therapy (the "shock and kill" strategy), has become a promising way to cure HIV-infected patients (3). Thus, understanding the HIV latency reactivation mechanism is vital and necessary for more effective drug target design using the "shock" strategy.

The main ingredients of the HIV regulatory loop are the long terminal repeat (LTR) and Tat transactivation on LTR. LTR is the promoter of the HIV genome and has a larger expression noise than promoters of human genes (4, 5). Nucleosomes associated with the LTR often block the full transcription by RNA polymerase (RNAP), resulting in a low basal expression rate. The rarely produced Tat protein complexes with CDK9 and CyclinT1 form a *positive transcriptional elongation factor b* (pTEFb). pTEFb can bind to the transactivation response element (TAR) the initially transcribed part of the HIV mRNA and remodel downstream nucleosomes. This remodeling assists elongating the mRNA, thus forming positive feedback (6-8). In addition, a bimodal gene expression ("phenotypic bifurcation" (9)) pattern was found in the offspring of defective-HIV infected cells with an initially intermediate expression (9). However, it was reported that the cooperativity coefficient (Hill coefficient) of Tat was only one (10), which means that the mean-field deterministic dynamics of HIV gene expression is monostable. This is distinct from the genetic toggle switch of the lambda phage regulatory loop with stronger feedback and bi-stable deterministic dynamics (11), or an oscillatory network with negative feedback and a limit cycle (12). The deterministic dynamics of the HIV regulatory network is insufficient to explain the observed bimodality, so a stochastic description may be required. Several stochastic models, such as two- or three-LTR-state models with or without positive feedback (13-17), have been proposed to study the dynamics of HIV gene expression. By combining the bimodality and noisiness of the HIV promoter gene expression, Weinberger et al. found that bimodality arises from a very slow rate of switching on LTR expression (9), resulting in much noisier dynamics than those observed for normal human promoters.

Recently, synergistic combinations of noise enhancers and activator drugs were reported not only to beat other reactivation cocktails in reactivating HIV latency but also to induce less cytotoxicity (14). Two specific types of drugs were involved in these synergistic combinations (14): Activators (AC), a small biological molecule that increases the average expression level of HIV proteins; and Noise Enhancers (NE), a different type of molecule that increases the noise of HIV protein expression but does not affect the average expression level. While NE itself cannot reactivate latent HIV, it was shown to amplify AC-induced reactivation of HIV significantly (14). The synergy gained from adding NE and AC to latent HIV is shown in Figure 1B. However, the biomolecular mechanisms underlying the synergy between AC and NE have not been fully resolved. Functions of only a small fraction of AC and NE are partially known. For example, as activators, TNF and Prostratin can activate the transcriptional factor NF-κB, and therefore antagonize HIV latency (18-20). Some of these Noise Enhancers, such as ethinyl estradiol, can influence HIV expression through another transcriptional factor SP1 or the structural state of chromatin (21, 22). The detailed molecular mechanisms of most noise enhancers are still unclear, indicating the complicated regulation mechanism of HIV dynamics. However, the noise enhancers in (14) only influence the system by regulating the transcription machinery, since the post-transcription noise-enhancing molecules are filtered out by a two-reporter assay.

One the other hand, thermodynamic energy dissipation and timescale play a crucial role in gene expression progress. A general model, which considers the binding of multiple transcription factors (TF) under thermodynamic equilibrium in prokaryotic cells and the function that different pair TF interactions can achieve in gene expression of cells, has already been studied extensively (23, 24). However, in studies of eukaryotic transcriptional dynamics, a nonequilibrium mechanism was found necessary (25), and many far-from-equilibrium models have been proposed (26-29). In addition to biomolecule synthesis and cell motility, the regulatory function of a living cell, such as adaptation, and the precise control of oscillations was also found highly dissipative (30, 31). Hence, we are very curious about whether certain energy input is necessary for noise modulated drug synergy in HIV latency reactivation. In addition, in a self-positive feedback gene regulatory network, the timescale of DNA state transition, mRNA transcription, mRNA decay, protein translation, and protein decay will influence the cell fate landscape and phenotype transitions (32-35). Post-integration HIV gene expression is one example system of the TF regulatory mechanism of gene expression with self-positive feedback. Hence, we are also interested in how timescales of gene-state transition and protein dynamics influence drug synergy.

To explore the mechanism of noise-enhanced drug synergy in reactivating latent HIV, we investigate an established LTR-two-state model and then propose a loop-LTR-four-state model that explains the noise enhanced drug synergy without direct interactions between AC and NE. Using these models, we then prove that, in the regulation of HIV promoter LTR, this synergy is controlled by the magnitude and direction of the system energy input. In essence, the LTR-four-state model is distinct from previous LTR state models, because the loop of the LTR four-state model allows for energy dissipation in the gene state transition network. Drug synergy can thus be significantly enhanced when we distribute the total energy input among two specific different reactions.

## Results

### NE and AC can exhibit drug synergy on reactivating HIV latency in the LTR-two-state model with significantly large $k_{off}$

We simulate the LTR-two-state model showed in Figure 2A using a Stochastic Simulation Algorithm (SSA) (36), calculating the reactivation of HIV latently infected T cells after adding NE and/or AC, with the appropriate assumptions on AC's and NE's functions (see Methods section).

Through simulation, we find that AC's inhibiting of NE's function on $k_{\text{on}}$ is necessary for synergy between AC and NE (Figure 2C–D). When AC has no inhibiting effect on NE, which corresponds to $f_{\text{inh}} = 0$, there is no synergy between AC and NE in the whole reasonable parameter space (Figure 2C; Figure 2B, black arrow). Only when AC inhibits NE's function of reducing $k_{\text{on}}$ can there be synergy between AC and NE in reactivating latent HIV (Figure 2D; Figure 2B, red arrow).

We also find that another necessary condition for NE to have significant synergy with AC is for $k_{\text{off}}$ to be no less than $10^{-1}$ $\text{hour}^{-1}$. However, as other important results have shown in the previous literature, the bimodal distribution of phenotype bifurcation (9) and HIV latency establishment operating autonomously from the host cellular state (16), are not sensitive to the increasing of $k_{\text{off}}$ (Figure S1G-I). When $k_{\text{off}}$ is less than $10^{-1}$, there is no synergy between AC and NE in reactivating latent HIV (Figure 2E; Figure S1E). However, if $k_{\text{off}}$ is no less than $10^{-1}$, then NE can have synergy with AC in reactivating latent HIV (Figure 2F–G; Figure S1F). Furthermore, the synergy between AC and NE will increase with the inhibiting effect quantity $f_{\text{inh}} > 0$ only when $k_{\text{off}}$ is sufficiently large, such as when $k_{\text{off}} = 0.8 \geq 10^{-1}$ (Figure 2G, red line). There will be no synergy between AC and NE with the inhibiting effect quantity $f_{\text{inh}} > 0$ if $k_{\text{off}} < 10^{-1}$ (Figure 2E, red line). Actually, in a latent HIV system where the LTR transcription-permissive state should be unstable, $k_{\text{off}}$ is more likely to be large due to the weak expression integration site.

In summary, we find two necessary conditions of drug synergy from our simulation results: (i) AC inhibits NE's function of reducing $k_{\mathrm{on}}$; (ii) there is a sufficiently large $k_{\mathrm{off}}$ (rate of LTR turning off or unbinding RNAP). However, the LTR-two-state model cannot uncover the mechanism of AC's inhibition on NE's function of reducing $k_{\mathrm{on}}$ , which is only an assumption made in (14). Furthermore, the very diverse NE selected from experiments also suggests that direct interaction between AC and the majority of NE is nearly impossible and a certain generic mechanism must be able to generate an equivalent effect.

**No drug synergy can be produced under a detailed balance condition in the LTR-four-state model coupling with Tat transactivation**

The two LTR states is oversimplified so that the condition (i) AC inhibits NE's function of reducing $k_{\mathrm{on}}$ for the noise-enhanced drug synergy based on the two-state model is not very natural and also is not convincing enough to explain why the drug synergy between activator and noise enhancer is almost universal as reported in (14). We implement a loop-LTR-four-state model coupled with a protein expression module (Figure 3A-C and 4A) to understand the mechanism of drug synergy in reactivating HIV latency. The rationale for choosing the four-state model rather than another other form is just from biochemistry. The activators used in the experiments mostly act on transcriptional factors which will bind the DNA in order to facilitate the binding of RNAP, so the DNA needs to bind both RNAP and transcriptional factors. The four-gene-state model (Figure 3A-C and 4A), in which each state represents a binding state of the promotor, is a widely-used description for modeling the promotor bounded both transcription factors (TF) and RNAP, just as in many previous studies on interactions between multiple TFs and RNAP in prokaryotic cells under thermodynamic equilibrium assumption (23, 24), or the model for investigating the allostery of two proteins through their binding with DNA (37). Thus, it is natural to chose this well-established four-state model for our study of drug synergy.

More specifically, for example, the activator TNFα can stimulate NF-κB, which can remodel the chromatin structure to become more RNAP-accessible (19, 38), hence in this situation, in the LTR-four-state model (Figure 3A, 4A), LTR state is exactly the transcription-inactive state without RNAP binding, corresponding to a restricted and inaccessible chromatin configuration (39); LTR* is the activated gene state without RNAP binding, corresponding to a more RNAP-accessible chromatin environment activated by host transcriptional factor, such as NF-κB; LTR-P and LTR*-P are the RNAP-binding states, respectively.

Similar as the well-analyzed assumptions of the AC's and NE's function from previous studies (14), AC is assumed to promote the transition of LTR or LTR-P to activated states, and NE is assumed to slowdown the RNAP binding/unbinding activities between the inactive gene states (Figure 3A and 4A). Nevertheless, in this LTR-four-state model, there is no assumption of any direct interaction between AC and NE. We divide the LTR-four-state model into two categories. One is with the detailed balance condition (Figure 3A), and the other, which will be studied in the next subsection, is without the detailed balance condition (Figure 4A).

Under the detailed balance condition (Figure 3A), our LTR-four-state model with the transcription/translation module without Tat transactivation (Figure 3B; see Methods section) illustrates that AC increases LTR mean expression level and that NE increases LTR expression noise (Figure S4C-D), which is consistent with the drug screening experimental results (Figure 1B) (14).

However, under the detailed balance condition, neither synergy between AC and NE nor depression of Noise Suppressor (NS) on AC, are possible in reactivating latent HIV (Figure 3). This contradicts the experimental data that shows that NE enhance AC's inducing of latent HIV reactivation or that NSs reduce AC's reactivating of latent HIV (Figure 1B) (14). More specifically, under the detailed balance condition, $P_{\mathrm{on}}$ stays the same when both NE (or NS) and AC are added to the system ($\gamma \gg 1$, $\alpha = 1$ (or $\alpha = -1$)), compared to when only AC is added ($\gamma \gg 1$, $\alpha = 0$) (Figure 3D; see Method Section 2.5 for details). Thus, the detailed-balanced LTR-four-state model predicts no synergy between NE and AC and predicts that NS does not suppress the AC's function of increasing $P_{\text{on}}$ (Figure 3D), contradicting the experimentally observed synergy between AC and NE (Figure 1B) (14). It is because under the detailed balance condition, the probability $P_{\text{on}}$ of RNA polymerase binding to LTR (LTR-P and LTR*-P) (see Method Section 2.1 and 2.5 for details) only depends on the equilibrium constants of each reaction, and NE or NS tunes the forward and backward rates simultaneously but keeps the equilibrium constant unvaried. This conclusion is not dependent on concrete models. It is a general physical and mathematical result. Hence it is not possible to build another even more complicated detailed-balanced model to overcome this obstacle.

We couple the detailed-balanced LTR-four-state model with Tat transactivation, and find there is still no synergy between NE and AC, as illustrated by the reactivation ratio of latent HIV (Figure 3E; see Method Section 2.7 for details). Note that the reactivation ratio of HIV is calculated dynamically for a finite time, starting from the latent state, which is different from its steady-state probability $P_{\text{on}}$. However, both are closely related to each other, since they both indicate the degree of reactivation for latent HIV.

In addition, we calculate the mean duration time (MDT) of both the LTR-off states (LTR and LTR*) and the LTR-on states (LTR-P and LTR*-P) (see Method Section 2.8 for details). The reciprocals of the MDTs calculated from the LTR-four-state model can be regarded as the effective transition rates in the reduced LTR-two-state model with only the LTR-off and LTR-on state. Next, we find out that AC can shorten the MDT at LTR-off states (Figure S4E; Figure S4F), and that NE can lengthen the MDT at both LTR-on states and LTR-off states with their ratio fixed (Figure 3F-G; S4G-H). These results are consistent with the assumptions of the LTR-two-state model in the previous section. However, in this detailed-balanced model, the effective inhibiting effect quantity $f_{\mathrm{inh}}$, as defined by the effective transition rates, always vanishes (Figure 3H; see Method Section 2.9 for the exact definition of $f_{\mathrm{inh}}$). This confirms the LTR-two-state model predictions that no synergy between AC and NE in reactivating latent HIV should be observed when $f_{\mathrm{inh}} = 0$.

Hence, for Noise Enhancers synergize with AC, the regulation of HIV gene expression must be a non-detailed-balanced process with energy dissipation.

**The direction of the cycle flux caused by energy input in the Non-Detailed-Balanced LTR-four-state model determines the synergy**

Inside a living cell, continuous energy consumption is necessary for executing different vital functions. We already know that systems with drug synergy must be energy dissipative, but how energy input, i.e., breaking the detailed balance, influences drug synergy remains poorly understood.

We mainly investigate how cycle flux direction and energy input distribution, as features of a nonequilibrium system, effect the synergy. Breaking the detailed balance is equivalent to having non-vanishing cycle fluxes. In our LTR-four-state model, the cycle fluxes can go either counter-clockwise or clockwise. Energy input can be distributed to one or more reactions. Here, we first consider the case of energy input for only one single reaction (Figure 4A; see Method Section 2.2 and Table S3 for details). In the real biological system, the energy input can be realized through ATP hydrolysis or other reversible covalent modification (40).

We prove that the non-detailed-balanced LTR-four-state model can produce the drug synergy between NE and AC on $P_{\mathrm{on}}$, if and only if the direction of cycle flux is clockwise. Mathematical analysis (see Method Section 2.10 for details) and numerical simulations illustrate the same phenomenon. The model with counter-clockwise cycle flux predicts no synergy between NE and AC on $P_{\mathrm{on}}$ or HIV latency reactivation, and no reduction of $P_{\mathrm{on}}$ or latent HIV reactivation is observed when NS is added with AC (Figure 4B). On the other hand, with a clockwise cycle flux, the model predicts in all cases that NE can synergize with AC on $P_{\mathrm{on}}$, and that 6 out of 8 cases NE synergize with AC on latent HIV reactivation (Figure 4C, Table S3). 4 out of 8 of the ways which break the detailed balance through a single reaction to produce a clockwise cyclic probability flux predicts that there is a significant synergy on $P_{\mathrm{on}}$ between NE and AC (Figure 4C up panel, Table S3), and that NS reduces $P_{\mathrm{on}}$ with AC added. Two out of 8 of the ways i.e., increasing the transition rate from LTR*-P to LTR-P, or reducing the transition rate from LTR-P to LTR*-P) predicts that there is a significant synergy on the reactivation of latent HIV between NE and AC (Figure 4C down panel, Table S3), and that NS reduces AC-induced HIV latency reactivation. The results of our model are consistent with the experimental fact that the majority of NE amplify AC reactivating latent HIV, while the majority of NSs suppresses reactivation of latent HIV with AC added. Thus, the non-detailed-balanced LTR-four-state model reveals that there is a general mechanism of the synergy between NE and AC on the reactivation of latent HIV, instead of a particular mechanism by a specific NE.

We also show that in the above cases of significant drug synergy between AC and NE, the clockwise cyclic probability flux always promotes LTR turn on mainly through the LTR*-to-LTR*-P pathway strengthened by AC and turn off through the LTR-P-to-LTR pathway weakened directly by NE (Figure 4D-E). It explains why NE can further amplify the HIV latency reactivation induced by AC, as long as the energy input provides clockwise cyclic probability flux.

In addition, for the equilibrium system, the probability density function of the dwell time at LTR-off states is predicted to be monotonically decreasing and convex (41) (Figure S10, solid black lines). Monotonicity or convexity can be maintained for the nonequilibrium system with a low magnitude of energy input (a small disturbance to the equilibrium system; see Figure S10, dashed red line). However, as the magnitude of energy dissipation increases, the nonmonotonicity or concavity of the probability density function of dwell time could appear (Figure S10D, H, dotted red line, and solid red line).

**The LTR-four-state model with distributed energy input may achieve much stronger synergy than that with energy input from a single reaction**

One possible strategy by which strong synergy can be achieved is to drive the LTR promoter to turn on mostly through the LTR-to-LTR*-to-LTR*-P pathway, whose rate can be significantly increased by AC, and to turn off mostly through the LTR*P-to-LTR-P-to-LTR pathway (Figure 4D-E), whose rate can be distinctly decreased by NE. This way, the promoter is more likely to transition to state LTR-P, rather than state LTR*, once it is at state LTR*-P. Here, we build an EITST (Energy Input on the Two Specific Transition rates) LTR-four-state model, in which part of the energy input reduces the transition rate from LTR*-P to LTR* ($\beta_2$) and the other part increases the transition rate from LTR*-P to LTR-P ($\beta_1$) (Figure 5A; see Method Section 2.2 for details), with the total energy fixed ($\beta_1 + \beta_2 = \beta$).

We find that there is an optimal energy input distribution ($\beta_1 = 1.8$, $\beta_2 = 8.2$) for the system to perform the strongest synergy between AC and NE on $P_{\mathrm{on}}$ (Figure 5B). The certain distributed energy input of $0 < \beta_1 < 10$ may achieve stronger synergy on $P_{\mathrm{on}}$ than that of a single reaction ($\beta_1 = 0$ or $\beta_1 = 10$). Drug synergy on HIV latency reactivation depends on an energy input distribution that will reach and then remain at the maximal level for $\beta_1 \gtrsim 4$ (Figure 5C). Without loss of generality, we set $\beta_1 = 5$ and $\beta_2 = 5$ for the EITST LTR-four-state model; all the following simulation results are based on these values.

In such a non-detailed-balanced model, simulation results of adding AC or NE alone with GFP present are consistent with the drug screening experimental data, that AC increases LTR mean expression level and NE increases LTR expression noise (Figure 5D-E, Figure 1B).

The synergy between a noise enhancer and an activator on both $P_{\mathrm{on}}$ (Figure 5G) and HIV latency reactivation (Figure 5H) have been observed to be much stronger than in the scenario where energy input is only from a single reaction. A noise enhancer can increase the HIV latency reactivation from approximately 7%, when an activator is already added, to 13%, when both are added (Figure 5H). These numbers are quite similar to the best cases observed in experiments when Prostratin is used as the AC (experimental data from Figure 3A in (14)). In addition, such a mechanism of noise-enhanced drug synergy is very robust (Figure S12B) when we replace the first-order degradation of protein Tat by a more realistic stochastic process in which the protein Tat continuously accumulates during a cell cycle and only halve upon cell division (Figure S12A) as suggested in previous studies (42-45). Further, a noise suppressor reduces the AC-induced HIV latency reactivation from 7% to less than 1% (Figure 5H). This synergy between AC and NE on both $P_{\mathrm{on}}$ and HIV latency reactivation is found to be positively correlated with the magnitude $|\beta|$ of the energy input, but reaches the maximum (for $P_{\mathrm{on}}$) and saturation (for HIV latency reactivation) when $|\beta|$ is sufficiently large ($\beta > 5$) (Figure S6A-B). In addition, the synergy is found to be positively correlated with the noise of NE (Figure S6C); this is consistent with the experimental data (Figure 3B in (14)).

We also calculate the mean duration time (MDT) of the LTR-off and LTR-on states. In contrast to the detailed-balanced situation, NE can lengthen MDT at LTR-on states more significantly than at LTR-off states (Figure 5F, Figure S7C-F). Further, the effective inhibiting parameter $f_{\mathrm{inh}} \approx 1 > 0$ (see Method Section 2.9 for the exact definition of this effective parameter) means that AC does inhibit NE's function of reducing the transition rate from LTR-off states to LTR-on states (Figure 5F). These simulation results verify the conclusion we made in the LTR-two-state model: that NE can synergize with AC and NE in reactivating latent HIV only when $f_{\mathrm{inh}} > 0$. Now we know that when AC inhibits NE's function of reducing effective $k_{\mathrm{on}}$, this is achieved by the energy input that drives the clockwise cycle flux.

However, as in the LTR-two-state model, a noise enhancer can amplify an activator's reactivating of latent HIV only if $k_{\mathrm{unbindp}}$ (equivalent to $k_{\mathrm{off}}$ in the LTR-two-state model) is greater than $10^{-2}$ (Figure S8A). To explain this necessary condition, we analyze the timescales of Tat transactivation dynamics and of LTR transitions. We find that it takes about $\tau_0 \approx 20$ hours on average of Tat transactivation for LTR to maintain an activated state for a long time (Figure S8B-C). Therefore, if $k_{\mathrm{unbindp}}$ is very small compared to the timescale of $1/\tau_0$ , the duration time of LTR-on state without NE present will be long enough for Tat transactivation to occur with a high probability; thus, further reducing $k_{\mathrm{unbindp}}$ by NE will have little effect (Figure S8B). When $k_{\mathrm{unbindp}}$ is not small compared to $1/\tau_0$, such a duration time is typically not long enough for Tat transactivation. In this case, lengthening the duration time of NE at the LTR-on state will provide Tat more time to reactivate latent HIV (Figure S8B), resulting in drug synergy with the activator.

Finally, to verify the model applicability, we use the same EITST model to explain other important previous experimental observations (Figure S3A-C), including Tat-transactivation-controlled HIV latency, which was establishedto operate autonomously from the host cellular state (16), and bimodal distribution in the phenotype bifurcation of the Tat level (9) (see Method Section 2.13 for parameter values). Also, in our EITST model, the nonmonotonicity and concavity of the probability density function's dwell time are observed to have a large magnitude of energy dissipation (Figure S10F).

**Discussion**

Long-lived latent HIV-1 is the main obstacle to a clinical cure (2). For noise-enhanced synergistic combinations of drugs that effectively reactivate HIV latency (14), we propose an LTR-four-state model with Tat transactivation (with only one cooperativity) to reveal the mechanism of this synergy, which is produced by the combination of AC and NE. Through analyzing and simulating this model, we find that the drug synergy on HIV latency reactivation depends on the distribution of energy input and the direction of the system's cycle flux.

As our model has illustrated, the synergy between AC and NE is universal, and our study supports the strategy of Dar et al. (2014) to discover novel drug combination for the treatment of virus infection, not only for HIV, but also for the virus with similar mechanism as HIV, such as the presence of latent state induced by the significant noise combined with weak positive feedback mechanism. The AC and NE identified for other virus can be different from those for HIV, but our model suggests that there should also been drug synergy between them.

Design principles for specific biological functions, such as reliable cell decisions (46), adaptation (47), robust and tunable biological oscillation (48), and the dual functions of adaptation and noise attenuation (49), have been extensively explored. Some of these functions, such as biochemical oscillations and adaptation, were found to depend on energy dissipation (30, 31). Here, we show that the drug synergy between NE and AC in reactivating HIV latency also depends on the direction which chemical energy is dissipated during the HIV LTR-state transition. This nonequilibrium property could also be used as a potential target for lentivirus latency reversal in synergetic therapeutic interventions. The optimization principle of energy input distribution for the highest drug synergy might also apply to network designing.

Our LTR-four-state model is a minimal model in which the effects of AC and NE are modeled to account for drug synergy. What we discover, through this generic model, is the presence of a generic mechanism that is not restricted to specific molecules. Without specifying the exact pathway of NE in the LTR expression, we here adopted the validated assumption of NE as proposed in the study by Dar et. al. (14). NE that are filtered to have no influence on post-transcription rates are assumed to simultaneously reduce the transition rate between LTR on states and off states (14). Actually, the non-transcription-permissive activated state of LTR (LTR*) in our model can represent different biochemical states of LTR in the process of HIV gene expression, depending on different AC and NE. The general mechanism of noise enhanced drug synergy we discovered is drawn from a rigorous mathematical proof (see Method section 2.10) and is valid within a 0.1-fold to 10-fold change to the parameters (see Figure S13 for sensitivity analysis). In addition, this LTR-four-state model can be expanded into a more detailed LTR-six-state model, where the Tat positive feedback is modeled through Tat binding to LTR and forms two new states, LTR-P-Tat and LTR*-P-Tat, as shown in a previous study (16). In the LTR-six-state model, the same synergy can be predicted (Figure S9, Method Section 2.4). Hence, the nonequilibrium mechanism of drug synergy we propose here is not dependent on a specific Tat positive feedback mechanism.

Most proteins are removed from the system primarily by dilution rather than active degradation mechanisms. We have shown that the mechanism for noise-enhanced drug synergy that we discovered is still valid if we replace the degradation process of protein with a noisier one due to partitioning at cell division. In real cells, the situation should be more complicated, for instance, the transcription is coupled with the cell size. However, what we have illustrated is the same mechanism works for the two extreme circumstances: one is the relatively smooth noise with first-order degradation if the transcription and cell size are perfectly coupled with each other, and the other is the partition of protein at the cell division after accumulation during a whole cell cycle without coupling to the cell size. Hence, we believe the same mechanism can be valid for the most general situations.

The nonequilibrium model proposed here is a minimal model providing an energy dissipation-based perspective to understand the general noise-enhanced drug synergy mechanism. In the LTR-two-state model, with or without Tat positive feedback, one necessary condition of the Noise-enhanced drug synergy in reactivating latent HIV is that AC blocks the NE's function of slowing down the rate of LTR when transitioning to a transcription-permissive state. However, this assumption cannot be justified or very well explained by previous two-state models. Improving upon earlier studies, the loop LTR-four-state model, along with a specific directional probability flux (caused by the energy dissipation), shows that LTR primarily turns on through the pathway strengthened by AC and turns off primarily through the pathway weakened by NE, resulting in a synergy between AC and NE when the dwelling time of LTR-transcription-permissive states is lengthened, and making latent HIV reactivation more likely. In the models without such transition loops, the synergy cannot emerge without assuming an interaction between AC and NE. In addition, the nonequilibrium LTR-four-state model in this article estimates the influence of the energy dissipation ($\beta \neq 0$) on the gene expressing process, including the mean duration time and the gene expression pattern, while in the LTR-two-state model, the energy dissipation cannot be modeled at all. This is because, only in the loop multi-gene-state-transition model, the detailed balanced condition can be violated through energy dissipation, while for a multi-state model without loops (e.g. a two-gene-state model), the detailed balanced condition with a steady distribution is inherently satisfied.

Our model predicts how the magnitude of energy dissipation influences the gene expression process of LTR and drug synergy, which is not possible for the two-state model. Recently, Wang, et al. (50) experimentally tuned phosphorylation energy in living cells and studied how it influences cell cycle dynamics. This kind of technique and experimental design could be applied to T-cells infected with LTR vectors in order to test the predicted relationship between drug synergy and energy dissipation in the LTR-four-state model.

## STAR★Methods

### LEAD CONTACT AND MATERIALS AVAILABILITY

Further information and requests for resources and reagents should be directed to and will be fulfilled by the Lead Contact, Hao Ge (haoge@pku.edu.cn).

### DATA AND CODE AVAILABILITY

**Simulations in this paper are performed in software MATLAB (https://www.mathworks.com/). The code is accessible here:**

https://github.com/Xiaolu-Guo/Noise-Enhanced-Drug-Synergy-in-HIV-Latency-Reactivation .

## METHOD DETAILS

### 1. LTR-2-State Model and Simulation

#### 1.1 LTR-2-state model

To investigate the synergy between AC and NE, we employed a well-established LTR-2-state model with Tat positive feedback from previous study (Figure 2A) (14, 16):

$$
\begin{gathered}
LTR_{\mathrm{on}} \xrightarrow{k_{off}} LTR_{\mathrm{off}} \\
LTR_{\mathrm{off}} \xrightarrow{k_{on}} LTR_{\mathrm{on}} \\
LTR_{\mathrm{on}} + Tat \xrightarrow{k_{bind}} LTR_{\mathrm{on-Tat}} \\
LTR_{\mathrm{on-Tat}} \xrightarrow{k_{unbind}} LTR_{\mathrm{on}} + Tat \\
LTR_{\mathrm{on}} \xrightarrow{k_{m}} mRNA + LTR_{\mathrm{on}} \\
LTR_{\mathrm{on-Tat}} \xrightarrow{k_{transact}} mRNA + LTR_{\mathrm{on-Tat}} \\
mRNA \xrightarrow{k_{p}} Tat + mRNA \\
mRNA \xrightarrow{d_{m}} \emptyset \\
Tat \xrightarrow{d_{p}} \emptyset
\end{gathered}
\qquad (S1)
$$

In this model, the promoter LTR can toggle between active and inactive states with transition rates $k_{off}$ and $k_{on}$. Tat can bind/unbind to LTR (TAR) with rate $k_{bind}$ and $k_{unbind}$, then transactivate LTR-on state once bound to TAR with a higher transcription rate $k_{transact}$ than the transcription rate $k_m$ at LTR-on state due to Tat's enhancing LTR transcriptional elongation. mRNA can translate into protein at rate $k_p$. Also, mRNA and Tat will degrade at rate $d_m$ and $d_p$ respectively.

#### 1.2 The functions of AC and NE in the LTR-2-state model

The values of parameters ($k_{\mathrm{on}}$, $k_{\mathrm{off}}$, $k_{\mathrm{transact}}$) are the same as those in (16), which is quantified by single-cell analysis (15, 51, 52). In eukaryotic cells, one of the major sources of gene expression noise is the burst transcription arising from the stochastic transition between active and inactive promoter states that correspond to closed or open chromatin states (53). Activators (AC) (e.g. Tumor Necrosis Factor (TNF)) can assist in the initiation of the transcript and then enhance the transcription frequency (15, 54). The burst frequency is determined by $\mathrm{k_{on}}$ in the LTR-two-state model. Thus, as dealt with in a previous study (14), AC is assumed to increase the parameter $\mathrm{k_{on}}$, where the changing ratio modeled by $\mathrm{r_{AC}}$. On the other hand, Dar, Hosmane (14) developed the two-reporter method (55-57) to filter noise enhancers not influenced by post transcription, and used noise analysis in a HIV system (15, 54), combined with experimental tests which infer that the effects of NE reduce $\mathrm{k_{on}}$ and $\mathrm{k_{off}}$ by the same ratio, enhancing the noise without changing the average expression of HIV. We here adopt the sound assumption that NE slow down the switching rates between the two LTR states (14), where the changing ratio is represented by $\mathrm{r_{NE}}$.

For clarity, we defined the rate variables as following:

$k_{on}(k_{off})$: the LTR turning on(off) rate in untreated HIV/LTR-GFP infected cells.

$k_{on,AC}(k_{off,AC})$: the LTR turning on(off) rate with only Activator added to HIV/LTR-GFP infected cells.

$k_{on,NE}(k_{off,NE})$: the LTR turning on(off) rate with only Noise Enhancer added to HIV/LTR-GFP infected cells.

$k_{on,AC,NE}(k_{off,AC,NE})$: the LTR turning on(off) rate with both Activator and Noise Enhancer added to HIV/LTR-GFP infected cells.

$k_{on,NS}$, $k_{off,NS}$, $k_{on,AC,NS}$ and $k_{off,AC,NS}$ are defined in the same way.

In this LTR-2-state model, the functions of AC and NE are assumed as following (14): adding AC increases $k_{on}$ to $k_{on}r_{AC}$, while adding NE reduces $k_{on}$ and $k_{off}$ to $k_{on}r_{NE}$ and $k_{off}r_{NE}$ ($r_{NE} < 1$), respectively, with their ratio fixed. From these assumptions, we have:

$$\begin{cases} k_{on,AC} = k_{on}r_{AC} \\ k_{off,AC} = k_{off} \end{cases}$$

and

$$\begin{cases} k_{on,NE} = k_{on}r_{NE} \\ k_{off,NE} = k_{off}r_{NE} \end{cases}$$

The values of $k_{on,AC,NE}$ and $k_{off,AC,NE}$ will be discussed in the next section.

### 1.3 $f_{inh}$, the degree of AC's inhibition upon the reduction of $k_{on}$ induced by NE

It is worth noting that Dar et al. mentioned, "Enhanced activation requires and assumes that any changes in $\mathrm{k_{on}}$ by the noise enhancer are overly-compensated by the activator" (4). In order to investigate whether this inhibition effect from the activator on the noise enhancer's function of changing $\mathrm{k_{on}}$ is necessary for drug synergy, we use the parameter $f_{\mathrm{inh}}$ to quantify the degree of AC's inhibition upon the NE-induced reduction of $k_{\mathrm{on}}$, as described following.

If the functions of AC and NE are working separately and independently, then

$$\begin{cases} k_{on,AC,NE} = k_{on} r_{AC} r_{NE} \\ k_{off,AC,NE} = k_{off} r_{NE} \end{cases}$$

However, there might be some interaction between AC and NE's function. For AC and NE to has synergy on HIV latency reactivation, AC might inhibit NE's function on $k_{on}$. We quantified the inhibition of AC on NE's function on $k_{on}$ as:

$$f_{inh} = 1 - \log_{r_{NE}}\left(\frac{k_{on,AC,NE}}{k_{on} r_{AC}}\right) = \frac{\ln(k_{on,AC,NE}) - \ln(k_{on,AC})}{\ln(k_{off,AC}) - \ln(k_{off,AC,NE})} + 1 \qquad \text{(S2)}$$

Then we have:

$$\begin{cases} k_{on,AC,NE} = k_{on} r_{AC} r_{NE}^{1-f_{inh}} \\ k_{off,AC,NE} = k_{off} r_{NE} \end{cases}$$

$$f_{inh} \in [0,1]$$

$f_{inh} = 0$ represents AC does not inhibit NE's function of reducing $k_{on}$ (Figure S1C, left panel), and $f_{inh} > 0$ means that AC does inhibit NE's function of reducing $k_{on}$. Particularly, $f_{inh} = 1$ means that NE's function of reducing $k_{on}$ is completely inhibited by AC (Figure S1C, right panel).

### 1.4 Simulation of reactivation ratio

The stochastic LTR-2-state model coupled with Tat positive feedback (Eqs (S1)) was simulated using the Stochastic Simulation Algorithm(SSA) (36).

Reactivation Ratio is the ratio of trajectory numbers with activated HIV (#Tat > 75) up to 100 hours and the total number of trajectories starting from the latency state ($LTR_{off} = 1$, copy numbers of all other species=0, simulated 5000~10000 cells) at time 0h. These simulations were implemented via Matlab™ with the parameters shown in Table S1 and Table S2.

### 1.5 The synergy between AC and NE

From the experiments, the drug synergy is that NE can significantly amplify the reactivation of latent HIV caused by AC, while NE itself cannot reactivate latent HIV, i.e. 1+1>2 (Figure 1B) (14). Mathematically, we define the synergy on reactivating latent HIV as:

$$\text{Synergy} = R_{AC,NE} - R_{AC}$$

$R_{AC,NE}$ is the reactivation ratio of the latent HIV under parameter $k_{on,AC,NE}$ and $k_{off,AC,NE}$, corresponding to adding AC and NE together; $R_{AC}$ is the reactivation ratio of the latent HIV under parameter $k_{on,AC}$ and $k_{off,AC}$, corresponding to adding only AC.

The calculation of reactivation ratio has been explained in section 1.4.

## 2. LTR-4-State Model and Simulation

### 2.1 Detailed Balance LTR-4-state model

We built a LTR-4-state model under detailed balance (Figure 3A):

$$
\begin{gathered}
LTR \xrightarrow{k_{act}\gamma} LTR^* \\
LTR^* \xrightarrow{k_{unact}} LTR \\
LTR^* \xrightarrow{\omega k_{bindp}} LTR^*P \\
LTR^*P \xrightarrow{k_{unbindp}} LTR^* \\
LTR^*P \xrightarrow{k_{unact}} LTRP \\
LTRP \xrightarrow{\omega k_{act}\gamma} LTR^*P \\
LTRP \xrightarrow{k_{unbindp}e^{-\alpha}} LTR \\
LTR \xrightarrow{k_{bindp}e^{-\alpha}} LTRP
\end{gathered}
\tag{S3}
$$

In our model, there are four different promotor states: LTR is the free state, LTR* is the activated state but without RNAP binding; LTR-P and LTR*-P are the corresponding RNAP-bounded states. AC is assumed to promote LTR transiting to the activated state LTR*, e.g. LTR bound with NF-κB, and LTR* recruits RNA polymerase much easier than LTR itself, e.g. the NF-κB bound to LTR acting as a Transcription Factor to recruit RNA polymerase to LTR(58). It has been shown that the screened Noise Enhancer has no effect on post transcription Dar, Hosmane (14), and some NE can increase transcription factors in cells, such as SP1 (21, 22). Similar to (14), we assume that NE, once present, can slows down the switching rates between LTR and LTR-P.

We used Markov jumping process to model the transition among LTR states with AC and/or NE added (Figure 3A). The four states can mutually transit. We assume $S = \{LTR, LTR^*, LTRP, LTR^*P\}$, use $R$ represents $LTR$ state, $R^*$ represents $LTR^*$ state, $R^*P$ represents $LTR^*P$ state, $P$ represents $LTRP$ state. Then $S = \{R, R^*, P, R^*P\}$. We denote that $ON = \{P, R^*P\}$, $OFF = \{R, R^*\}$. The generator matrix (transition rate matrix) is:

$$Q=\begin{bmatrix} k_R & k_{R,R^*} & k_{R,RP} & 0 \\ k_{R^*,R} & k_{R^*} & 0 & k_{R^*,R^*P} \\ k_{P,R} & 0 & k_P & k_{P,R^*P} \\ 0 & k_{R^*P,R^*} & k_{R^*P,P} & k_{R^*P} \end{bmatrix} \tag{S4}$$

$$k_i = -\sum_j k_{i,j} \quad i,j \in S$$

Then we can calculate invariant distribution $\pi$:

$$\begin{bmatrix} \pi_R \\ \pi_{R^*} \\ \pi_P \\ \pi_{R^*P} \end{bmatrix}^T \begin{bmatrix} k_R & k_{R,R^*} & k_{R,P} & 0 \\ k_{R^*,R} & k_{R^*} & 0 & k_{R^*,R^*P} \\ k_{P,R} & 0 & k_P & k_{P,R^*P} \\ 0 & k_{R^*P,R^*} & k_{R^*P,P} & k_{R^*P} \end{bmatrix} = \begin{bmatrix} 0 \\ 0 \\ 0 \\ 0 \end{bmatrix}^T \tag{S5}$$

More specifically, here we assume in the absence of AC, RNAP binds to LTR at a relatively slow rate $k_{bindp}[P]$ and unbinds fast at rate $k_{unbindp}$; LTR transit to LTR* state with an extremely slow rate $k_{act}$ without AC, but at a much higher rate $k_{act}\gamma$ ($\gamma \gg 1$) once AC is present; RNAP is attracted to bind to LTR* at a higher rate $\omega k_{bindp}$, where $\omega$ is the cooperative interaction factor ($\omega > 1$); when NE is added, LTR will bind and unbind RNAP at a slower rate ($k_{bindp}e^{-\alpha}$, $k_{unbindp}e^{-\alpha}$) with a reduction parameter $\alpha > 0$; LTR-P and LTR*-P can mutually transit at rate $\omega k_{act}\gamma$ and rate $k_{unact}$.

Here, there is no external energy input; it is under detailed balance condition:

$$k_{R,R^*}k_{R^*,R^*P}k_{R^*P,P}k_{P,R} = k_{R,P}k_{P,R^*P}k_{R^*P,R^*}k_{R^*,R}$$

where $k_{R,R^*} = k_{act}\gamma$, $k_{R^*,R^*P} = \omega k_{bindp}$, $k_{R^*P,P} = k_{unact}$, $k_{P,R} = k_{unbindp}e^{-\alpha}$, $k_{R,P} = k_{bindp}e^{-\alpha}$, $k_{P,R^*P} = \omega k_{act}\gamma$, $k_{R^*P,R^*} = k_{unbindp}$, $k_{R^*,R} = k_{unact}$.

## 2.2 Non-Detailed-Balance LTR-4-state model

Breaking the detailed balance condition in the Detailed Balance LTR-4-state model, we can build non-Detailed-Balance LTR-4-state models. We first built it with energy input only through a single transition: Energy input can influence any single transition rate in the Detailed Balance LTR-4-state model through multiplying it by a factor of $e^{\beta}$. Such an energy input will cause clockwise (c.w.) probability flux or counter-clockwise (c.c.w.) probability flux. The details of the non-Detailed-Balance model are listed in Table S4.

We also built an EITST(Energy Input on the Two Specific Transition rates) LTR-4-state model, in which part of the energy input is through reducing the transition rate from LTR*-P to LTR* by multiplying $e^{-\beta_2}$, and the other half is through increasing the transition rate from LTR*-P to LTR-P by multiplying $e^{\beta_1}$ (Figure 6A), instead of with energy input on a single transition. The details of the EITST model are shown below.

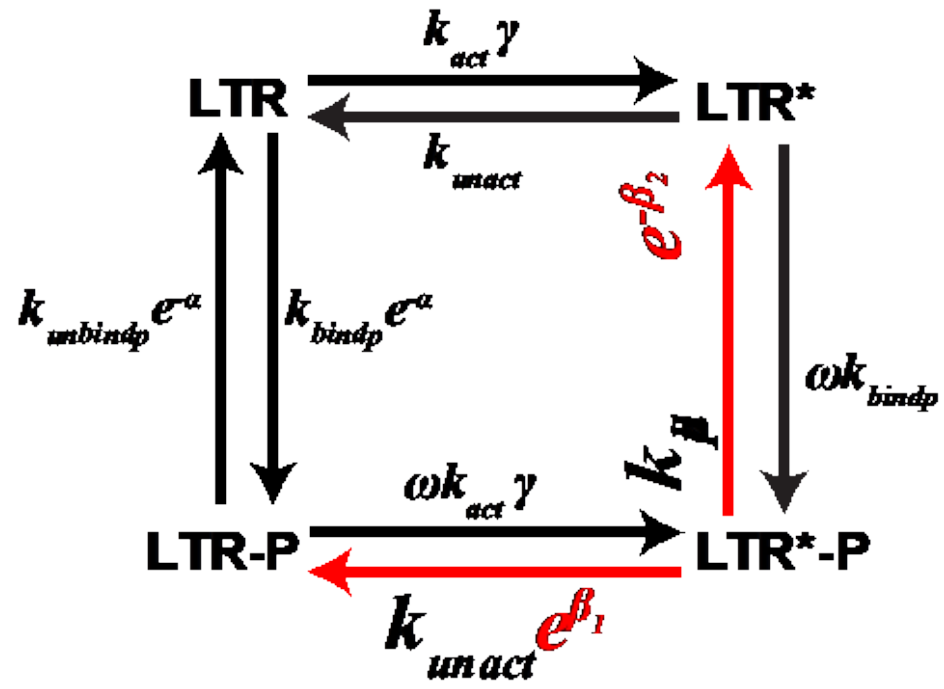

Figure. The rate formula in EITST model

## 2.3 GFP expression without feedback, calculation of mean and noise of LTR

We describe the dynamics of LTR-GFP vector expression using the following chemical reactions (combined with reactions (S3)):

$$LTR^*P \xrightarrow{k_m} mRNA + LTR^*P$$

$$LTRP \xrightarrow{k_m} mRNA + LTRP$$

$$mRNA \xrightarrow{k_p} GFP + mRNA \tag{S6}$$

$$mRNA \xrightarrow{d_m} \emptyset$$

$$GFP \xrightarrow{d_p} \emptyset$$

For the LTR-GFP vector, with RNAP bond to LTR (i.e. LTR-P state or LTR*-P state), the downstream DNA of LTR can be transcribed into mRNA at rate $k_m$ and then translate into protein at rate $k_{GFP}$. Also, mRNA and GFP will degrade at rate $d_m$ and $d_{GFP}$ respectively (Figure 3B). To Calculate the Noise and Mean of GFP of this system, we need calculate the first and second moment of GFP:

$$< GFP > (t) = \sum_{i\in S} \sum_{m=0}^{\infty} \sum_{n=0}^{\infty} nP(LTR = i, mRNA = m, GFP = n, t)$$

and

$$< GFP^2 > (t) = \sum_{i\in S} \sum_{m=0}^{\infty} \sum_{n=0}^{\infty} n^2 P(LTR = i, mRNA = m, GFP = n, t)$$

Here, $P(LTR = i, mRNA = m, t)$ represents the probability of LTR staying at $i$ state and $\#mRNA = m$ at time $t$; $P(LTR = i, mRNA = m, GFP = n)$ represents the probability of LTR staying at $i$ state and $\#mRNA = m$ and $\#GFP = n$ at time $t$. For the following, when the variable/quantity/moment is not written as an explicit function of time $t$, it mean the steady state value. Then we sum up the related master equation and calculate the steady state (the derivative is zero):

$$< GFP >= \frac{k_p}{d_p} < mRNA > \tag{S7}$$

and

$$< GFP^2 >= \frac{k_p}{d_p}(< GFP\ mRNA > + < mRNA >) \tag{S8}$$

To calculate the above quantity, we need calculate the following moments of mRNA and GFP:

$$< mRNA > (t) = \sum_{i \in S} \sum_{m=0}^{\infty} mP(LTR = i, mRNA = m, t)$$

$$< mRNA^2 > (t) = \sum_{i \in S} \sum_{m=0}^{\infty} m^2 P(LTR = i, mRNA = m, t)$$

$$< GFP\ mRNA > (t) = \sum_{i \in S} \sum_{m=0}^{\infty} \sum_{n=0}^{\infty} mnP(LTR = i, mRNA = m, GFP = n, t)$$

and similarly, we sum up the related master equation and calculate the steady state (the derivative is zero), followed by

$$< mRNA >= \frac{1}{d_m} \sum_{i \in S} k_{m,i} \pi_i \tag{S9}$$

$$< GFP\ mRNA >= \frac{\sum_{i \in S} k_{m,i} < GFP >_i + k_p < mRNA^2 >}{d_m + d_p} \tag{S10}$$

$$< mRNA^2 >= \frac{1}{d_m} \sum_{i \in S} k_{m,i} < mRNA >_i + \frac{1}{d_m} \sum_{i \in S} k_{m,i} \pi_i \tag{S11}$$

where

$$k_{m,i} = \begin{cases} k_m, i \in ON \\ 0, i \in OFF \end{cases}$$

$$< mRNA >_i = \sum_{m=0}^{\infty} mP(LTR = i, mRNA = m)$$

$$< GFP >_i = \sum_{n=0}^{\infty} n \sum_{m=0}^{\infty} P(LTR = i, mRNA = m, GFP = n)$$

for all $i \in S$. $< mRNA >_i$ and $< GFP >_i$ satisfy the linear equations (the steady state of Master Equations):

$$(d_m I_4 - Q)\begin{bmatrix} < mRNA >_R \\ < mRNA >_{R^*} \\ < mRNA >_P \\ < mRNA >_{R^*P} \end{bmatrix} = \begin{bmatrix} k_{m,R}\pi_R \\ k_{m,R^*}\pi_{R^*} \\ k_{m,P}\pi_P \\ k_{m,R^*P}\pi_{R^*P} \end{bmatrix} \tag{S12}$$

$$(d_p I_4 - Q)\begin{bmatrix} < GFP >_R \\ < GFP >_{R^*} \\ < GFP >_P \\ < GFP >_{R^*P} \end{bmatrix} = k_p \begin{bmatrix} < mRNA >_R \\ < mRNA >_{R^*} \\ < mRNA >_P \\ < mRNA >_{R^*P} \end{bmatrix} \tag{S13}$$

where $I_4$ is the $4 \times 4$ identity matrix, $Q$ is the generator matrix for the LTR-4-state model.

We solved the linear equations (S12) (S13), then substituted $< mRNA >_i$ and $< GFP >_i$ for $i \in S$ into equations (S10) (S11). We then substituted (S9) (S10) (S11) into equations (S7) (S8). Using the above calculation and submission, we have the Noise of LTR-GFP vector, $\frac{< GFP^2 > - < GFP >^2}{< GFP >^2}$, and the mean of LTR-GFP vector, $< GFP >$.

## 2.4 Tat expression with positive feedback

We describe the dynamics of the full length HIV vector expression with Tat positive feedback using the following chemical reactions (combined with reactions (S3)):

$$\begin{gathered} LTR^*P \xrightarrow{k_m} mRNA + LTR^*P \\ LTRP \xrightarrow{k_m} mRNA + LTRP \\ mRNA \xrightarrow{k_{Tat}} Tat + mRNA \\ mRNA \xrightarrow{d_m} \emptyset \\ Tat \xrightarrow{d_{Tat}} \emptyset \end{gathered} \tag{S14}$$

The Tat forms positive feedback by enhancing the elongation of initial transcribed mRNA of HIV (59, 60) and by stabilizing the HIV activation (16). We model these two functions by following:

$$k_m = k_{mbasal} + k_{trs1}\frac{\frac{Tat}{k_{trs2}}}{1+\frac{Tat}{k_{trs2}}} \tag{S15}$$

and

$$k_{unbindp} = \frac{K_{threshold}^3 + \delta Tat^3}{K_{threshold}^3 + Tat^3} k_{unbindp0} \tag{S16}$$

All parameter values are shown in Table S5.

To prove the model results does not depend on Tat active degradation, we also performed the simulation with cell division. Similar as previous works simulating cell division (61), in our corresponding SSA, every time the updated time passes one cell cycle length, the system will execute cell division, where Tat concentration will be diluted to half of its concentration. The simulation results shown in Figure S12.

## 2.5 Probability of LTR-on states $P_{on}$

From (S5), we have invariant distribution of LTR-4-state model, $\pi_i$ for $i \in S$. We then calculated the probability of LTR-on states (LTR-P state and LTR*-P state):

$$P_{on} = \pi_P + \pi_{R^*P}$$

## 2.6 Probability flux and cycle flux

From invariant distribution (S5) and the transition rates, we can calculate the probability flux of LTR-4-state model at steady state:

$$J_{ij} = \pi_i k_{i,j}$$

for $i, j \in S$. And net flux from state I to state j is defined as $J_{ij} - J_{ji}$.

In such a 4-state model, there is only one cycle (LTR->LTR*->LTR*-P->LTR-P->LTR), which is clockwise, and its reversed one. The net cycle flux of the clockwise cycle is $J_c = J_{R,R*} = J_{R*,R*P} = J_{R*P,RP} = J_{RP,R}$, and the net flux of the reversed counterclockwise cycle is $-J_c$.

## 2.7 Reactivation ratio

The stochastic LTR-4-state model coupled with Tat positive feedback (Eqs (S3) (S14) (S15) (S16)) was simulated using the Stochastic Simulation Algorithm(SSA), or say 'Gillespie' algorithm (36), because of the difficulty to analytically calculate the model with feedback.

The reactivation ratio is the ratio of activated HIV (#Tat > 75), where the trajectory number is 100 hours and the total number of trajectories starts from the latent state (LTR=1, copy numbers of all other species=0, simulated 5000~10000 cells) at time 0h. These simulations were implemented via Matlab™ with the parameters shown in Table S4 and Table S5.

## 2.8 Mean duration time

We need a theorem to calculate mean duration time.

**Theorem** (62)

Let $\{X_t, t \geq 0\}$ be a continuous time Markov chain on state space $S$, with generator matrix $Q = (q_{ij})$. $S_1$ and $S_2$ are subspaces of $S$ satisfying:

$$S = S_1 \cup S_2,\ S_1 \cap S_2 = \emptyset,\ S_1 \neq \emptyset,\ S_2 \neq \emptyset.$$

Suppose the invariant distribution $\mu = \{\mu_i : i \in S\}$ exists, then the mean duration in $S_1$ (denoted by $\tau$) takes the form of

$$\tau = \frac{\sum_{i \in S_1} \mu_i}{\sum_{i \in S_1} \sum_{j \in S_2} \mu_i q_{ij}}.$$

See (62) for proof.

For the LTR-4-state continuous time Markov chain with transitions showed in Eqs (S3), the generator $Q$ is (S4), and the invariant distribution $\pi_i$ for $i \in S$ can be derived from linear Eqs (S5). We assume that $S_1 = ON = \{LTRP, LTR^*P\}$, $S_2 = OFF = \{LTR, LTR^*\}$. We define $\tau_{ON}(\tau_{OFF})$ as the mean duration time of LTR stay at ON(OFF) states. By the theorem, we have:

$$\tau_{ON} = \frac{\sum_{i \in ON} \pi_i}{\sum_{i \in ON} \sum_{j \in OFF} \pi_i q_{ij}}$$

$$\tau_{OFF} = \frac{\sum_{i \in OFF} \pi_i}{\sum_{i \in OFF} \sum_{j \in ON} \pi_i q_{ij}}$$

## 2.9 $f_{inh}$, the degree of AC's inhibition upon the reduction of $\lambda_{on}$ induced by NE

We regarded the reciprocal of the Mean Duration Time as the transition rates between LTR-on and LTR-off states, $\lambda_{on}$ and $\lambda_{off}$,

$$\begin{cases} \lambda_{on} = \dfrac{1}{\tau_{OFF}} \\ \lambda_{off} = \dfrac{1}{\tau_{ON}} \end{cases} \tag{S17}$$

which are equivalent to $k_{on}$ and $k_{off}$ in the effective LTR-2-State model. Then we defined the effective $f_{inh}$ using the formula (S2), i.e.

$$f_{inh} = \frac{\ln(\lambda_{on,AC,NE}) - \ln(\lambda_{on,AC})}{\ln(\lambda_{off,AC}) - \ln(\lambda_{off,AC,NE})} + 1 \tag{S18}$$

$\lambda_{on,AC}$ and $\lambda_{off,AC}$ correspond to the LTR-state model with only AC added, i.e. $\gamma \gg 1$, $\alpha = 0$; $\lambda_{on,AC,NE}$ and $\lambda_{off,AC,NE}$ correspond to the LTR-state model with both AC and NE added, i.e. $\gamma \gg 1$, $\alpha > 0$. Similar we can define $\lambda_{on,AC,NS}$ and $\lambda_{off,AC,NS}$.

## 2.10 Theorem on the relation between drug synergy of $P_{on}$ and cyclic probability flux

Generally, our non-detailed-balanced LTR-4-state model can be described as below:

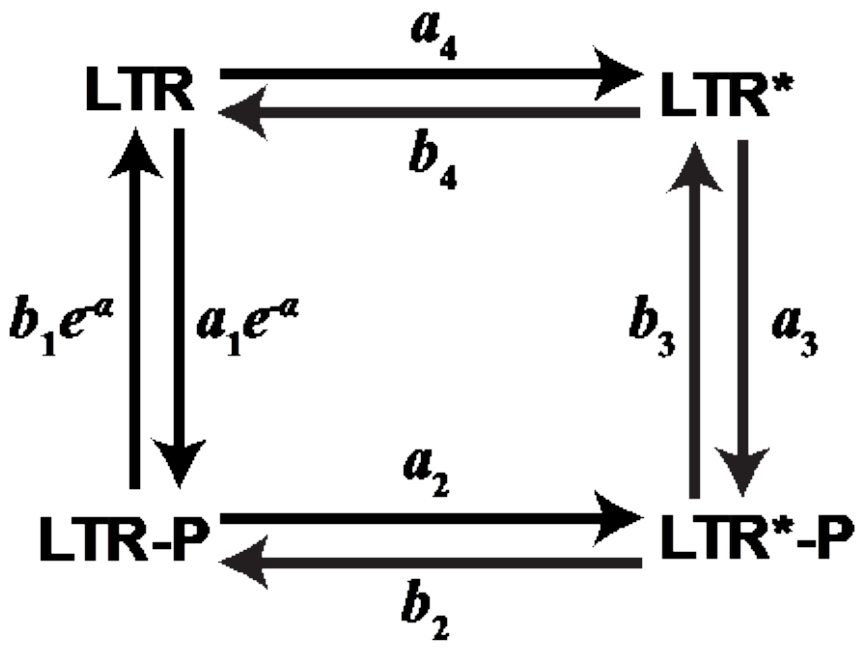

The corresponding Chemical Master Equations are

$$\begin{cases} \frac{dP_{LTR}}{dt} = -(a_1 e^{-\alpha} + a_4)P_{LTR} + b_1 e^{-\alpha} P_{LTRP} + b_4 P_{LTR^*} \\ \frac{dP_{LTRP}}{dt} = a_1 e^{-\alpha} P_{LTR} - (b_1 e^{-\alpha} + a_2) P_{LTRP} + b_2 P_{LTR^*P} \\ \frac{dP_{LTR^*P}}{dt} = a_2 P_{LTRP} - (b_2 + b_3) P_{LTR^*P} + a_3 P_{LTR^*} \\ \frac{dP_{LTR^*}}{dt} = a_4 P_{LTR} + b_3 P_{LTR^*P} - (b_4 + a_3) P_{LTR^*} \end{cases}$$

At steady state, the derivative is zero. Then, with the net flux $J$ introduced, the equations above at steady state can be simplified to a set of equations:

$$\begin{cases} J = -a_1 e^{-\alpha} P_{LTR} + b_1 e^{-\alpha} P_{LTRP} \\ J = -a_2 P_{LTRP} + b_2 P_{LTR^*P} \\ J = -b_3 P_{LTR^*P} + a_3 P_{LTR^*} \\ J = -b_4 P_{LTR^*} + a_4 P_{LTR} \end{cases} \tag{S19}$$

Here, $J > 0$ indicates a clockwise cycle net flux while $J < 0$ means a counter-clockwise one. And note that **synergy** of $P_{on}$ is defined as the increase of $P_{on}(\alpha) = P_{LTRP}(\alpha) + P_{LTR^*P}(\alpha)$ when there exists NE ($\alpha > 0$) and the decrease of $P_{on}(\alpha)$ when NS is present ($\alpha < 0$), compared with $P_{on}$ in the merely-AC case ($\alpha = 0$)

**Theorem 1.** For any positive $a_i$ and $b_i$ $(i = 1, 2, 3, 4)$,

(1) If $J > 0$, then $P_{on}(\alpha) > P_{on}(0)$ for $\alpha > 0$ and $P_{on}(\alpha) < P_{on}(0)$ for $\alpha < 0$;

(2) If $J < 0$, then $P_{on}(\alpha) < P_{on}(0)$ for $\alpha > 0$ and $P_{on}(\alpha) > P_{on}(0)$ for $\alpha < 0$.

**Proof.** We begin with the case of $\alpha = 0$. From the kinetics equations (S19), we can eliminate the probabilities at the OFF state:

$$P_{LTR} = -\frac{J}{a_1} + \frac{b_1}{a_1} P_{LTRP},$$

$$P_{LTR^*} = -\frac{a_1 + a_4}{a_1 b_4} J + \frac{a_4 b_1}{a_1 b_4} P_{LTRP}.$$

Therefore,

$$\left(1 + \frac{a_3(a_1 + a_4)}{a_1 b_4}\right) J = -b_3 P_{LTR^*P} + \frac{a_3 a_4 b_1}{a_1 b_4} P_{LTRP},$$

$$J = b_2 P_{LTR^*P} - a_2 P_{LTRP},$$

and

$$P_{LTRP} = \frac{a_1 b_4 (b_2 + b_3) + a_3 b_2 (a_1 + a_4)}{a_3 a_4 b_1 b_2 - a_1 a_2 b_3 b_4} J,$$
$$P_{LTR^*P} = \frac{a_1 a_2 (a_3 + b_4) + a_3 a_4 (a_2 + b_1)}{a_3 a_4 b_1 b_2 - a_1 a_2 b_3 b_4} J.$$

Since $P_{LTR} + P_{LTR^*} + P_{LTRP} + P_{LTR^*P} = 1$ gives that

$$-\frac{J}{a_1} + \frac{b_1}{a_1} P_{LTRP} - \frac{a_1 + a_4}{a_1 b_4} J + \frac{a_4 b_1}{a_1 b_4} P_{LTRP} + P_{LTRP} + P_{LTR^*P} = 1,$$

we have

$$-\frac{a_1 + a_4 + b_4}{a_1 b_4} + \frac{a_4 b_1 + b_1 b_4 + a_1 b_4}{a_1 b_4} \frac{a_1 b_4 (b_2 + b_3) + a_3 b_2 (a_1 + a_4)}{a_3 a_4 b_1 b_2 - a_1 a_2 b_3 b_4} + \frac{a_1 a_2 (a_3 + b_4) + a_3 a_4 (a_2 + b_1)}{a_3 a_4 b_1 b_2 - a_1 a_2 b_3 b_4} = \frac{1}{J}.$$

By replacing $\mathrm{a}_1$ and $b_1$ with $\mathrm{a}_1 e^{-\alpha}$ and $b_1 e^{-\alpha}$, we obtain the general equality:

$$\begin{aligned} & a_1 (a_2 a_3 + a_2 b_3 + a_2 b_4 + a_3 b_2 + b_2 b_4 + b_3 b_4) \\ & \quad + e^{\alpha} (a_2 a_3 a_4 + a_2 a_4 b_3 + a_2 b_3 b_4 + a_3 a_4 b_2) \\ & \quad + b_1 (a_3 a_4 + a_3 b_2 + a_4 b_2 + a_4 b_3 + b_2 b_4 + b_3 b_4) \\ & \quad = \frac{a_3 a_4 b_1 b_2 - a_1 a_2 b_3 b_4}{J(\alpha)}. \end{aligned}$$

It is easy to see that:

(1) $J(\alpha)$ decreases with $\alpha$ when $J > 0$ and increases with $\alpha$ when $J < 0$, i.e.

$$J(0) - J(\alpha) > 0, \text{ if } J > 0 \qquad \text{(S20)}$$
$$J(0) - J(\alpha) < 0, \text{ if } J < 0$$

(2) $e^{\alpha} J(\alpha)$ increases with $\alpha$ when $J > 0$ and decreases with $\alpha$ when $J < 0$;

$$e^{\alpha} J(\alpha) - J(0) > 0, \text{ if } J > 0 \qquad \text{(S21)}$$
$$e^{\alpha} J(\alpha) - J(0) < 0, \text{ if } J < 0$$

(3) $J(\alpha)$ shares the same sign with $J(0)$.

$$J(\alpha) J(0) > 0$$

When NE/NS concentration approaches zero which means $\alpha = 0$, our model reduces to the general equality in Jia et al. (63):

$$P_{\mathrm{on}}(\alpha) - P_{\mathrm{on}}^0 = \left(P_{\mathrm{on}}^{\infty} - P_{\mathrm{on}}^0\right) P_{\mathrm{on}}(\alpha) - \left(\frac{P_{\mathrm{on}}^{\infty}}{a_3} - \frac{P_{\mathrm{on}}^0}{a_1}\right) J, \qquad \text{(S22)}$$

where $P_{\mathrm{on}}^0 = \frac{a_1}{a_1 + b_1}, P_{\mathrm{on}}^{\infty} = \frac{a_3}{a_3 + b_3}$ remain the same under any value of α. From (S22), we have:

$$\left(1 - P_{\mathrm{on}}^{\infty} + P_{\mathrm{on}}^0\right) P_{\mathrm{on}}(\alpha) = P_{\mathrm{on}}^0 - \left(\frac{P_{\mathrm{on}}^{\infty}}{a_3} - \frac{P_{\mathrm{on}}^0}{a_1}\right) J, \qquad \text{(S23)}$$

To calculate the synergy $P_{on}(\alpha) - P_{on}(0)$, we subtract equation (S23) with $\alpha = 0$ from equation (S23) with $\alpha > 0$ (NE), and we have:

$$\left(1 - P_{\text{on}}^{\infty} + P_{\text{on}}^{0}\right)\left(P_{\text{on}}(\alpha) - P_{\text{on}}(0)\right) = -\left(\frac{1}{a_3+b_3} - \frac{e^{\alpha}}{a_1+b_1}\right)J(\alpha) + \left(\frac{1}{a_3+b_3} - \frac{1}{a_1+b_1}\right)J(0)$$

$$= \frac{J(0)-J(\alpha)}{a_3+b_3} + \frac{e^{\alpha}J(\alpha)-J(0)}{a_1+b_1}. \quad \text{(S24)}$$

Since $1 - P_{\text{on}}^{\infty} + P_{\text{on}}^{0} > 0$, and for clockwise cyclic probability flux ($J > 0$), substitute (S20) (S21) into (S24), we can prove that such case predicts positive synergy.

Similarly, for clockwise cyclic probability flux ($J > 0$), for the effect of NS ($\alpha < 0$),

$$\left(1 - P_{\text{on}}^{\infty} + P_{\text{on}}^{0}\right)\left(P_{\text{on}}(\alpha) - P_{\text{on}}(0)\right) = \frac{J(0) - J(\alpha)}{a_3 + b_3} + \frac{e^{\alpha}J(\alpha) - J(0)}{a_1 + b_1} < 0$$

which shows that the LTR-4-state model with clockwise cyclic probability flux distinguishes NE and NS well.

Since the monotonicity of both $J(\alpha)$ and $e^{\alpha}J(\alpha)$ gets reverse, by the same token we can prove that for counter-clockwise cyclic probability flux ($J < 0$), $P_{\text{on}}(\alpha) - P_{\text{on}}(0) < 0$ and no synergy is predicted when $\alpha > 0$ (NE), while $P_{\text{on}}(\alpha) - P_{\text{on}}(0) > 0$ when $\alpha > 0$ (NS).

## 2.11 Distribution of duration time at LTR-on/off states in equilibrium system

According to Tu (41), the distribution of duration time at LTR-on/off states in equilibrium system should be monotonically decreasing and convex, while in the non-equilibrium system these features can be violated. We use the following equations to calculate the distribution of duration time at LTR-off states:

$$\frac{dQ(LTR,\tau)}{d\tau} = k_{R^*,R}Q(LTR^*,\tau) - k_{R,R^*}Q(LTR,\tau) - k_{R,P}Q(LTR,\tau)$$

$$\frac{dQ(LTR^*,\tau)}{d\tau} = -k_{R^*,R}Q(LTR^*,\tau) + k_{R,R^*}Q(LTR,\tau) - k_{R^*,R^*P}Q(LTR^*,\tau)$$

With initial conditions:

$$Q(LTR,0) = A_{off}k_{P,R}\pi_P$$

$$Q(LTR^*,0) = A_{off}k_{R^*P,R^*}\pi_{R^*P}$$

Where

$$A_{off} = \frac{1}{k_{P,R}\pi_P + k_{R^*P,R^*}\pi_{R^*P}}$$

The distribution of duration time at LTR-off states can be expressed:

$$P_{off}(\tau) = k_{R,P}Q(LTR,\tau) + k_{R^*,R^*P}Q(LTR^*,\tau)$$

### 3. Using Tat-binding states to model Tat positive feedback

To prove the results is independent from the specific details of the model, we also applied another model adapted from (16) shown below.

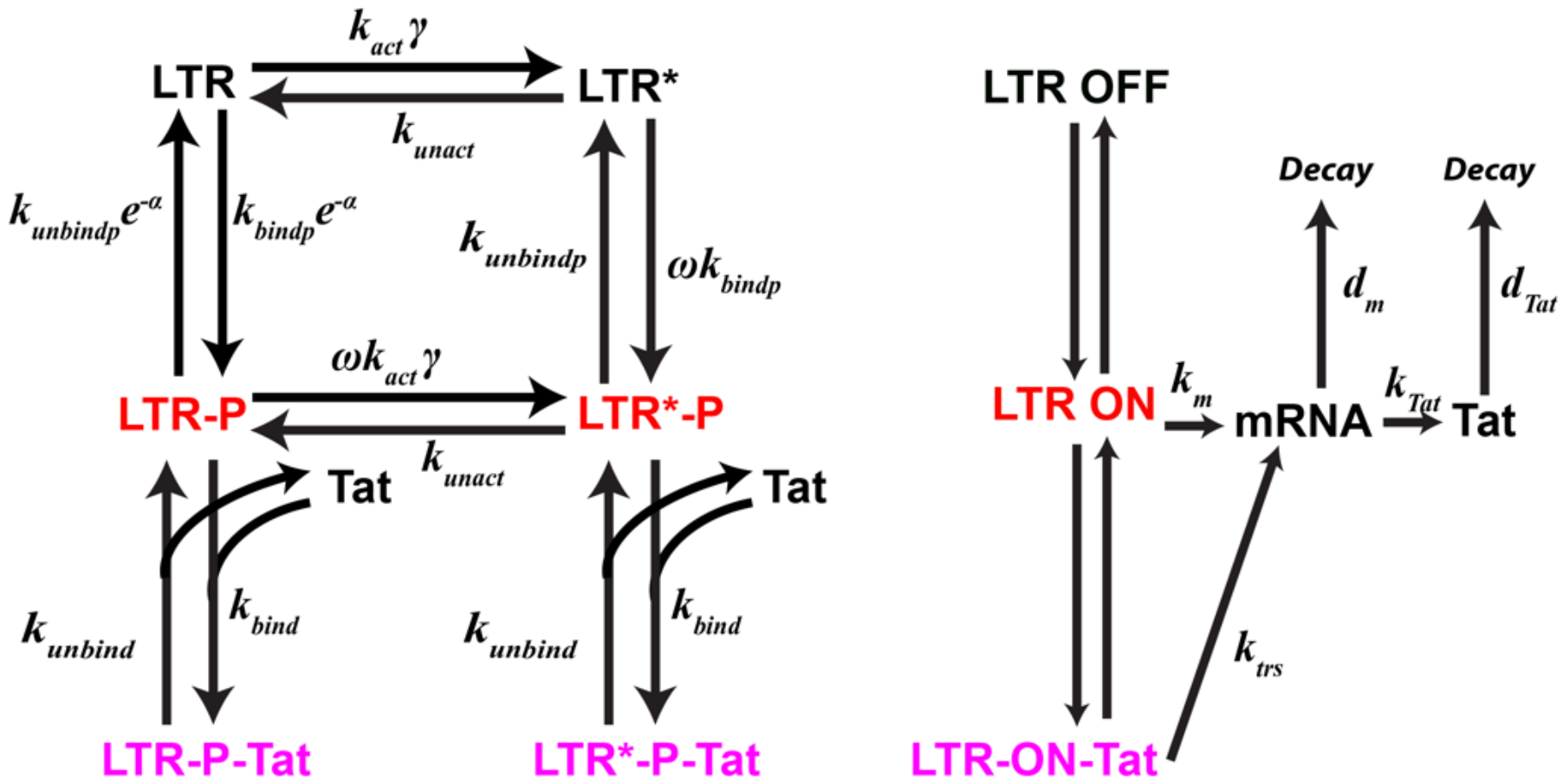

In this model, similar to LTR-2-state Tat positive feedback model, the Tat forms positive feedback through binding to LTR(TAR) Tat can bind/unbind to LTR(TAR) with rate $\boldsymbol{k_{bind}}$ and $\boldsymbol{k_{unbind}}$, then transactivate LTR-on state once bound to TAR with a much higher transcription rate $\boldsymbol{k_{trs}}$ than the transcription rate $\boldsymbol{k_m}$ at LTR-on state due to Tat's enhancing LTR transcriptional elongation. The system can be expressed by the following chemical reactions (combined with reactions (S3)):

$$\begin{gathered}
\boldsymbol{LTRP + Tat \xrightarrow{k_{bind}} LTRP_{Tat}} \\
\boldsymbol{LTRP_{Tat} \xrightarrow{k_{unbind}} LTRP + Tat} \\
\boldsymbol{LTR^{*}P + Tat \xrightarrow{k_{bind}} LTR^{*}P_{Tat}} \\
\boldsymbol{LTR^{*}P_{Tat} \xrightarrow{k_{unbind}} LTR^{*}P + Tat} \\
\boldsymbol{LTRP \xrightarrow{k_m} mRNA + LTRP} \\
\boldsymbol{LTR^{*}P \xrightarrow{k_m} mRNA + LTR^{*}P} \\
\boldsymbol{LTRP_{Tat} \xrightarrow{k_{trs}} mRNA + LTRP_{Tat}} \\
\boldsymbol{LTR^{*}P_{Tat} \xrightarrow{k_{trs}} mRNA + LTR^{*}P_{Tat}} \\
\boldsymbol{mRNA \xrightarrow{k_{Tat}} Tat + mRNA} \\
\boldsymbol{mRNA \xrightarrow{d_m} \emptyset}
\end{gathered} \tag{S25}$$

$$\boldsymbol{Tat} \xrightarrow{d_{Tat}} \emptyset$$

All parameter values are the same as LTR-2-state Tat positive feedback model, shown in Table S1.

## Author contributions

H.G. and L.Z. conceived and supervised the project; X.G. performed the majority of data analysis and mathematical modeling with contributions by T.T.; X.G. performed the majority of analysis and computational simulations with contributions by M.D.; X.G., H.G. and L.Z. wrote the paper.

## Acknowledgments

We would like to thank Professors Yuhai Tu and Tom Chou for helpful discussions. We would like to thank Dr. Lu Klinger's contribution to edit and polish the paper.

This work was supported by the National Natural Science Foundation of China (12050002, 11622101, 11971037) and the National Key R&D Program of China 2021YFF1200500.

## Authors' Conflicts of Interest

The authors declare no competing interests.

**Figures**

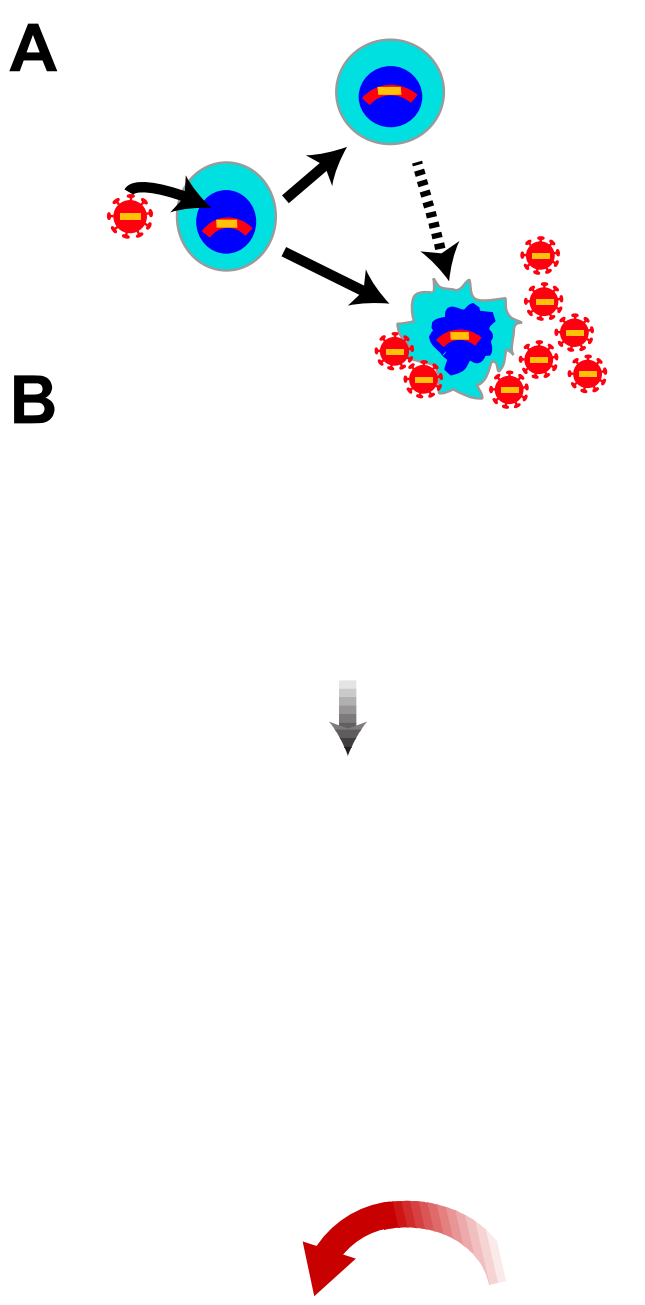

**Figure 1**. **HIV-infected cell fates and biological function of biomolecules reactivating latent HIV.**

**(A)** Schematic of different fates of cells when infected by HIV: HIV active replication, HIV proviral latency, and HIV latency reactivation (adopted from Figure 1A of (14)).

**(B)** Diagram of screening Activators (AC) and Noise Enhancers (NE) (up) and testing synergy on the reactivation of latent HIV after adding AC or/and NE (down). In previous experiments, AC and NE were selected by detecting the mean and noise of LTR expression using cells infected by the LTR-GFP vector. The synergy between AC and NE on HIV reactivation was tested using cells infected by full-length HIV with Tat transactivation. Untreated cells (grey bar) represent a control group. In comparison to the control group, adding the Activator (green bar) increases LTR expression, while adding the Noise Enhancer (magenta bar) increases LTR noise. Adding AC and NE simultaneously (red bar) has a strong synergy on HIV reactivation (which increases reactivation of latent HIV infected Cells). Adding AC and noise suppressors (NS, blue bar) has a depressing effect on HIV latency reactivation compared to adding AC only. (14)

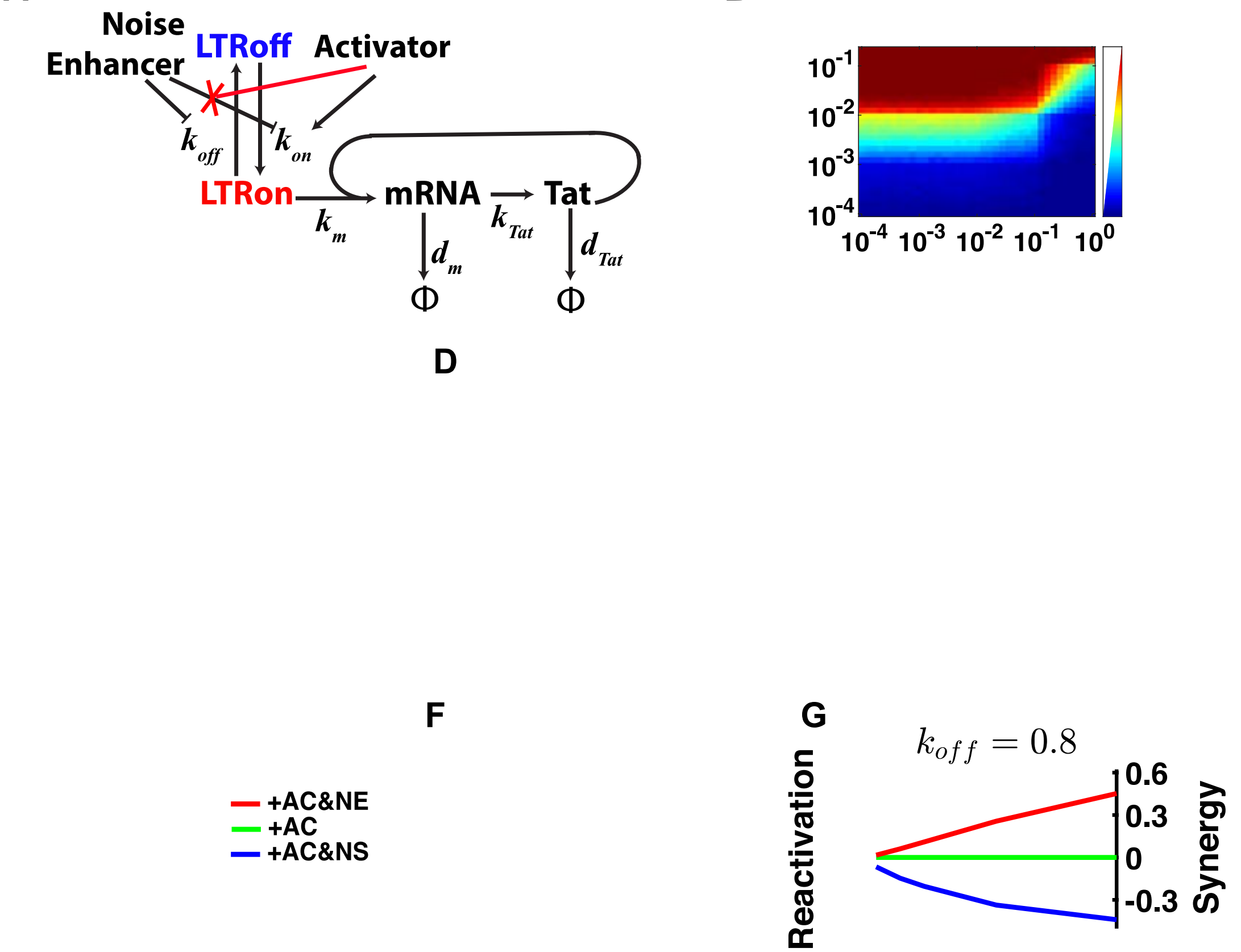

**Figure 2. Two necessary conditions for drug synergy**

**(A)** Modified from Figure 3A of (16). The LTR-two-state model with Tat feedback is used to explain the effects of NE and AC molecules on HIV. LTR has two states, on and off, which convert to each other at the rate of $k_{\mathrm{on}}$ and $k_{\mathrm{off}}$; the LTR-on state transcribes HIV mRNA at a rate of $k_{\mathrm{m}}$, mRNA degrades at a rate of $d_{\mathrm{m}}$ rate or translates to protein at a rate of $k_{\mathrm{Tat}}$, and Tat degrades at a rate of $d_{\mathrm{Tat}}$. NE decreases $k_{\mathrm{on}}$ and $k_{\mathrm{off}}$ with their ratio fixed; AC molecules will increase $k_{\mathrm{on}}$; AC inhibits NE's function on $k_{\mathrm{on}}$ when added together as assumed in (14). Tat transactivates LTR through enhancing the transcriptional rate $k_{\mathrm{trasact}}$.

**(B)** The heat map of reactivation across different values of LTR turning off at rate $k_{\mathrm{off}}$ and LTR turning on at rate $k_{\mathrm{on}}$. The green arrow corresponds to adding AC. The red arrow corresponds to adding NE but only decreasing $k_{\mathrm{off}}$ without changing $k_{on}$, i.e., AC inhibits NE's function on $k_{\mathrm{on}}$. The black arrow corresponds to adding NE, decreasing $k_{\mathrm{off}}$ and $k_{\mathrm{on}}$ with their ratio fixed, i.e., AC does not inhibit NE decreasing $k_{\mathrm{on}}$. (See Table S1 for parameter values.)

**(C-D)** The heat map of synergy on reactivation without AC's inhibition on NE ($f_{\mathrm{inh}} = 0$) and with AC's complete inhibition on NE ($f_{\mathrm{inh}} = 1$), respectively, across different values of $k_{\mathrm{off}}$ and $k_{\mathrm{on}}$, with AC added. The synergy is the reactivation plus NE and AC, subtracting the reactivation where only AC is added. (See Table S1 for parameter values.)

**(E-G)** The plots of synergy on reactivation have different $f_{inh}$ values for small $k_{\mathrm{off}}$, intermediate $k_{\mathrm{off}}$, and large $k_{\mathrm{off}}$, respectively. The red lines indicate adding NE and AC simultaneously; the green lines indicate adding AC only; and the blue lines indicate adding NS and AC simultaneously. (See Table S1 and Table S2 for parameter values.)

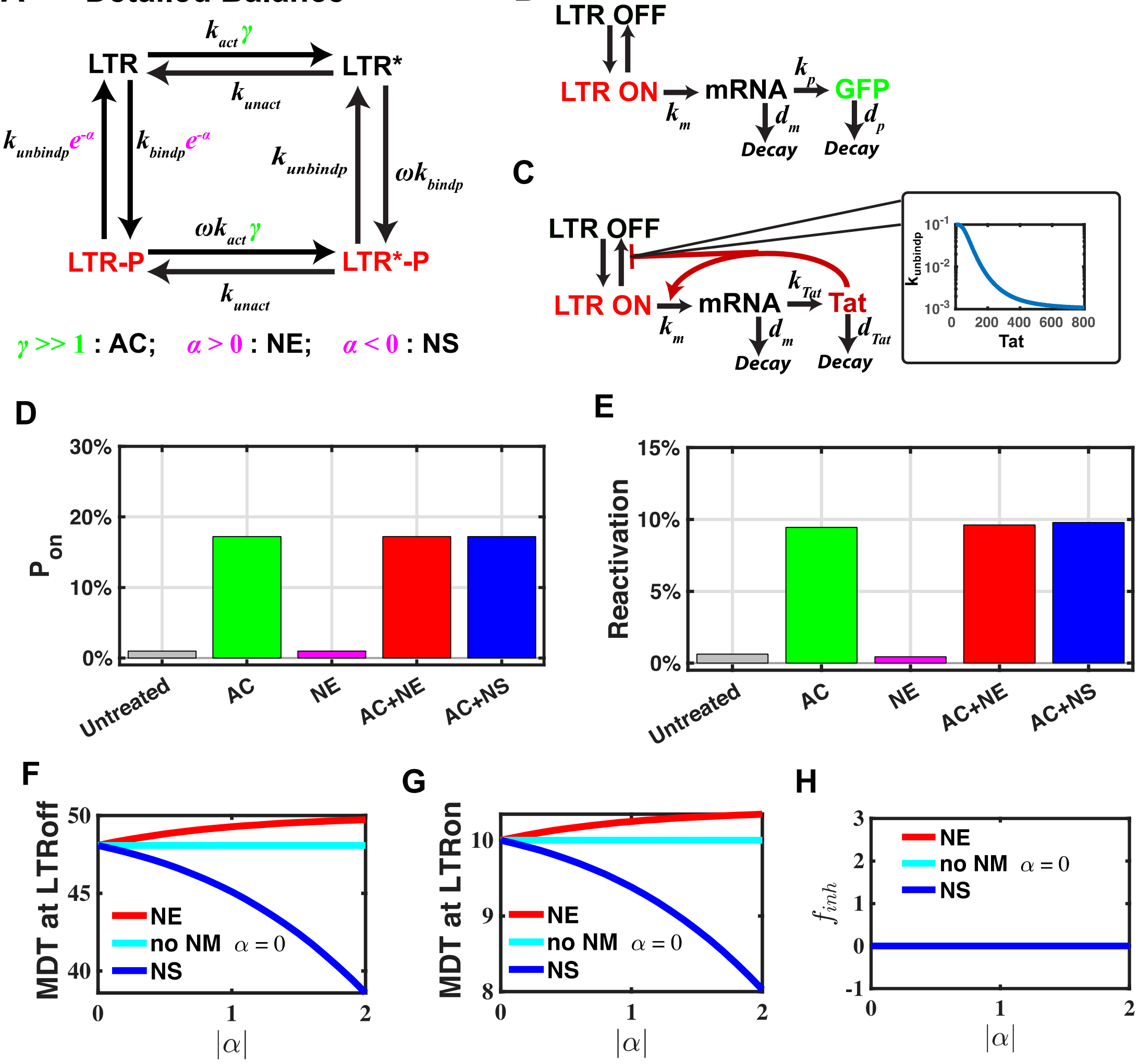

**Figure 3**. **No synergy is predicted under the detailed balance condition**

**(A)** Schematic of the LTR-four-state model with the Tat-feedback circuit. The LTR promoter is modeled for four states: a transcriptional silence state (LTR state), in which there are extremely slow binding RNAP polymerase or activation transcription factors such as NF-κB; an activated state (LTR*), such as LTR with a NF-κB bond; a transcription-permissive state (LTR-P); a transcription-permissive state with NF-κB bond (LTR*-P). Here, $k_{\mathrm{act}}$ is the rate for LTR binding NF-κB, while $k_{\mathrm{unact}}$ is the rate for LTR unbinding NF-κB. $\gamma$ models AC as the rate for LTR binding NF-κB increases. $\gamma = 1$ for untreated HIV infected cells, and $\gamma \gg 1$ when adding AC. $\omega$ is the attraction coefficient between NF-κB and RNAP, $\omega = 10$ ($\omega > 1$ means NF-κB attracts RNAP); $k_{\mathrm{bindp}}$ is the rate at which RNAP binds to LTR; $k_{\mathrm{unbindp}}$ is the rate at which RNAP unbinds from LTR; $\alpha$ is the noise attenuation factor ($\alpha > 0$ corresponds to the noise enhancer, and $\alpha < 0$ corresponds to noise suppressor). The parameters set here follow the detailed balance condition. The case of breaking the detailed balance can be seen in Figure 4 and Table S3.

**(B)** Schematic of the LTR-four-state model coupled with the transcription and translation module without feedback. LTR-on states (red, including LTR-P state and LTR*-P state) transcribes mRNA at rate $k_{\mathrm{m}}$; mRNA decays at rate $d_{\mathrm{m}}$ or can be translated at rate $k_{\mathrm{p}}$ into GFP; GFP decays at rate $d_{\mathrm{p}}$.

**(C)** Schematic of the LTR-four-state model coupled with the transcription and translation module with the Tat-transactivation circuit. LTR-on states (red, including LTR-P state and LTR*-P state) transcribes mRNA at rate $k_{\mathrm{m}}$; mRNA decays at rate $d_{\mathrm{m}}$ or can be translated at rate $k_{\mathrm{Tat}}$ into Tat; Tat decays at rate $d_{\mathrm{Tat}}$; Tat has positive feedback on $k_{\mathrm{m}}$; Tat stabilizes the state of LTR-on state through negative feedback on $k_{\mathrm{unbindp}}$.

**(D-E)** Probability of LTR-on states (LTR-P state andLTR*-P state), $P_{\mathrm{on}}$, and reactivation ratio of Latent HIV, respectively, in the detailed-balanced LTR-four-state model. Y-axis is the $P_{\mathrm{on}}$ and the reactivation ratio value, respectively, and X-axis is the categories of different combinations of AC and NE/NS. Untreated cases (grey bars) correspond to $\gamma = 1$, $\alpha = 0$; adding only AC (green bars) corresponds to $\gamma \gg 1$, $\alpha = 0$; adding only NE (magenta bars) corresponds to $\gamma = 1$, $\alpha = 1$; adding NE and AC (red bars) corresponds to $\gamma \gg 1$, $\alpha = 1$; adding NS and AC (blue bars) corresponds to $\gamma \gg 1$, $\alpha = -1$. (E) We use the Stochastic Simulation Algorithm (SSA) to calculate the reactivation ratio of the LTR-four-state model coupled with Tat feedback. The reactivation ratio is the ratio of the reactivated HIV trajectory number at time 100h to all trajectory numbers, starting from the latency state (LTR=1, all other species=0, simulated 5000 cells).

**(F-G)** Mean duration time at LTR-off states and LTR-on states, respectively, under the detailed balance condition.

**(H)** $f_{\mathrm{inh}}$ is the degree of AC's inhibition upon the reduction of $k_{\mathrm{on}}$ induced by NE under the detailed balance condition. We first calculate the reciprocal of the mean duration time as the transition rate between LTR-on and LTR-off states, $\lambda_{\mathrm{on}}$ and $\lambda_{\mathrm{off}}$, respectively. Then we calculate using the formula $f_{inh} = \frac{\ln(\lambda_{on,AC,NE})-\ln(\lambda_{on,AC})}{\ln(\lambda_{off,AC})-\ln(\lambda_{off,AC,NE})} + 1$ (see equations S17 and S18 for more details). (F-H) The red lines correspond to adding AC and NE ($\alpha > 0$); the cyan lines correspond to adding AC only ($\alpha = 0$); the blue lines correspond to adding AC and NS ($\alpha < 0$). (See Table S5 and Table S6 for parameter values.)

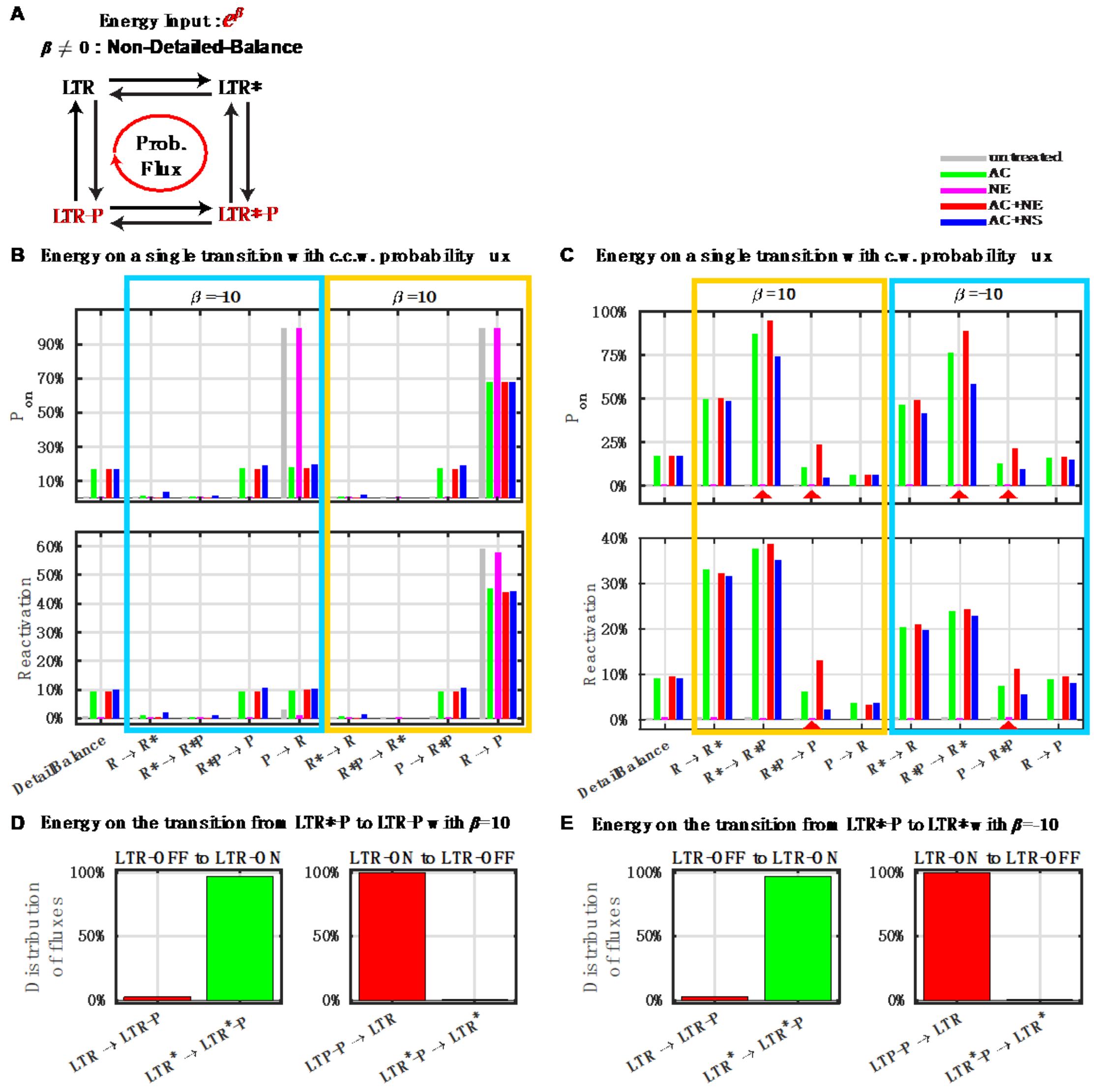

**Figure 4. The LTR-four-state model with energy input on a single transition produces synergy between AC and NE if and only if the system has clockwise cyclic probability flux**

**(A)** Schematic of the non-detailed-balanced LTR-four-state model. Energy input can influence any single transition rate of the detailed-balanced LTR-four-state model with the corresponding rate multiplying by $e^{\beta}$ or $e^{-\beta}$. Such an energy input will cause clockwise (c.w.) cyclic probability flux or counter-clockwise (c.c.w.) cyclic probability flux.

**(B-C)** Probability of LTR-on states, $P_{\text{on}}$ (up panels), and reactivation ratio of latent HIV (down panels) calculated from the non-detailed-balanced models with energy input on different single transitions causing a c.c.w. cyclic probability flux (B) or c.w. cyclic probability flux (C). R stands for LTR state; R* stands for LTR* state; P stands for LTR-P state; R*P stands for LTR*-P state. Each group of x-axis represents the non-detailed-balanced model with the corresponding transition rate multiplying by $e^{\beta}$ ($\beta > 0$ for orange groups, $\beta < 0$ for blue groups). Red triangles indicate the significant synergy cases. (See Table S3 for precise values plotted) (See Table S3 for model details. See Table S5 and Table S6 for parameter values.)

**(D-E)** The distributions of fluxes for LTR turning on (left panels) and turning off (right panels).

(See Table S3 and Method Section 2.4-8 for model details. See Table S5 and Table S6 for parameter values.)

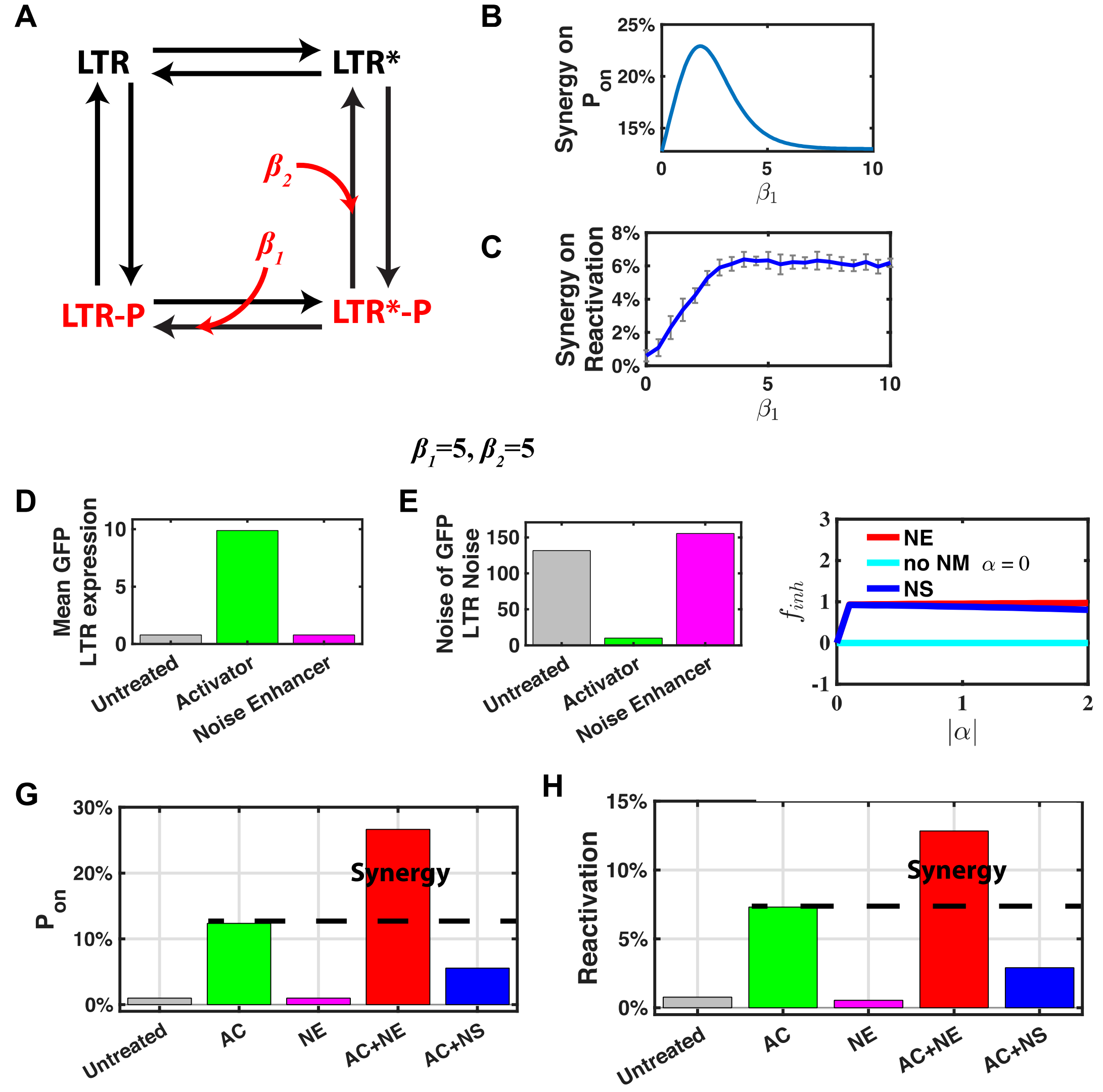

**Figure 5. LTR-four-state model with distributed energy input exhibits strong synergy between AC and NE**

**(A)** Schematic of the EITST model with distributed energy input ($\beta = \beta_1 + \beta_2 = 10$). The first part of the energy $\beta_1$ increases the LTR*-P-to-LTR-P transition rate by multiplying $e^{\beta_1}$; the other part of the energy $\beta_2$ reduces the LTR*-P-to-LTR* transition rate by multiplying $e^{-\beta_2}$.

**(B-C)** The synergy on $P_{\mathrm{on}}$ (B) and HIV latency reactivation (C) varies with energy distribution ($\beta_1 + \beta_2 = \beta = 10$). (C) Each point has an average of 250 simulation experimental data points, with 10000 cells simulated for each experiment. Error bars show the standard deviation.

**(D-J)** $\beta_1 = \beta_2 = 5$.

**(D-E)** The mean and noise of GFP expression calculated from the EITST model without positive feedback.

**(F)** $f_{\mathrm{inh}}$ is the degree of AC's inhibition upon the reduction of $k_{\mathrm{on}}$ induced by NE of the EITST model.

**(G-H)** Probability of LTR-on states (LTR-P state and LTR*-P state), $P_{\mathrm{on}}$, and reactivation ratio of latent HIV, calculated from the EITST model. (See Table S3 for model details. See Table S5 and S6 for parameter values.)